\documentclass[12pt,reqno]{lms}  

\usepackage{strings}
\usepackage{cite}

\title[{A tube formula for the Koch snowflake curve.}]
  {A tube formula for the Koch snowflake curve, with applications to complex dimensions.}
\author[M. L. Lapidus and E. P. J. Pearse]{Michel L. Lapidus and Erin P. J. Pearse}

\date{\today.}

  \classno{26A30, 28A12, 28A75, 28A80 (primary); 11K55, 26A27, 28A78, 28D20, 42A16 (secondary)}

 \extraline{The work of MLL was partially supported by the US National Science Foundation under grant DMS-0070497. Support by the Centre Emile Borel of the Institut Henri Poincar\'{e} (IHP) in Paris and by the Institut des Hautes \'{E}tudes Scientifiques (IHES) in Bures-sur-Yvette, France, during completion of this article is also gratefully acknowledged by MLL.}

\numberwithin{equation}{section} \numberwithin{theorem}{section}

\begin{document}

\newcommand{\draught}{false} \newcommand{\rough}{false}
\newcommand{\qed}{\makebox[0.1ex]{\;} \hfill \ensuremath \square}

\maketitle

\begin{abstract}
  A formula for the interior \ge-neighbourhood of the classical von Koch
  snowflake curve is computed in detail. This function of \ge is
  shown to match quite closely with earlier predictions
  from \cite{La-vF1}
  of what it should be, but is also much more precise. The
  resulting `tube formula' is expressed in terms of the Fourier
  coefficients of a suitable nonlinear and periodic analogue of
  the standard Cantor staircase function and reflects the
  self-similarity of the Koch curve. As a consequence, the
  \possible complex dimensions of the Koch
  snowflake are computed explicitly.
\end{abstract}

\allowdisplaybreaks

\section{Introduction}
  \label{sec:introduction}
  In \cite{La-vF1}, the authors lay the foundations for a theory
  of complex dimensions with a rather thorough investigation of
  the theory of fractal strings (see also, e.g.,
  [\citen{LaPo1}--2,\citen{LaMa},\citen{La1}--3,\citen{HaLa},\citen{HeLa},\citen{La-vF2}--3]);
  that is, fractal subsets of \bR.
  Such an object may be represented by a sequence of bounded open
  intervals of length $l_j$:
  \begin{equation} \label{notation:fractal-string}
    \L := \{l_j\}_{j=1}^\iy, \q \text{with } \sum_{j=1}^\iy l_j < \iy.
  \end{equation}
  The authors are able to relate geometric and physical properties
  of such objects through the use of zeta functions which contain
  geometric and spectral information about the given string. This
  information includes the dimension and measurability of the
  fractal under consideration, which we now recall.

  For a nonempty bounded open set $\gW \ci \bR$, $V(\ge)$ is defined
  to be the \emph{inner \makebox{\ge-neighborhood}} of \gW:
  \begin{equation} \label{notation:V-of-eps}
    V(\ge):= \vol \{x \in \gW \suth d(x,\del \gW) < \ge\},
  \end{equation}
  where \vol denotes 1-dimensional Lebesgue measure.
  Then the \emph{Minkowski dimension} of the boundary \del\gW
  (i.e., of the fractal string \L$\mspace{-10mu}$ ) is
  \begin{equation} \label{notation:minkowski-dimension}
    D=D_{\del\gW}=\inf\{t \geq 0 \suth V(\ge)
     =O\left(\ge^{1-t}\right) \text{ as } \ge \to 0^+\}.
  \end{equation}
  Finally, \del\gW is \emph{Minkowski measurable} if and only if the
  limit
  \begin{equation} \label{notation:minkowski-measurable}
    \M = \M(D;\del\gW)
       = \lim_{\ge \to 0+} V(\ge) \ge^{-(1-D)}
  \end{equation}
  exists, and lies in $(0,\iy)$. In this case, \M is called the
  Minkowski content of $\del\gW$.

  If, more generally, \gW is an open subset of $\bR^d$,
  then analogous \defns hold if $1$ is replaced
  by $d$ in  (\ref{notation:V-of-eps})--(\ref{notation:minkowski-measurable}).
  Hence, $\vol[d]$ denotes the $d$-dimensional volume (which is area for $d=2$)
  in the counterpart of  (\ref{notation:V-of-eps}) and $\ge^{1-t},
  \ge^{-(1-D)}$ are replaced by $\ge^{d-t}$ in (\ref{notation:minkowski-dimension}),
  and $\ge^{-(d-D)}$ in (\ref{notation:minkowski-measurable}),
  \resp. In (\ref{notation:V-of-eps})--(\ref{notation:minkowski-measurable}),
  $d=1$ and the positive numbers $l_j$ are the
  lengths of the connected components (open intervals) of \gW,
  written in nonincreasing order. In much of the rest of the
  paper (where \gW is the snowflake domain of $\bR^2$), we have
  $d=2$. See, e.g.,
  [\citen{Man},\citen{Tr},\citen{La1},
   \citen{LaPo1}--2,\citen{Mat,La-vF1}]
  and the relevant references therein for further information on
  the notions of Minkowski--Bouligand dimension (also called `box
  dimension') and Minkowski content.

  The complex dimensions of a fractal string \L are defined to be the
  poles (of the \mero \cntn) of its \emph{geometric zeta function}
  \begin{equation} \label{notation:geom-zeta-fn}
    \gz_\L(s) = \sum_{j=1}^\iy l_j^s,
  \end{equation}
  in accordance with the result that
  \begin{equation} \label{notation:minkowski-dimension-of-a-string}
    D_\L = \inf \{\gs \geq 0 \suth \sum_{j=1}^\iy l_j^\gs < \iy\},
  \end{equation}
  i.e., that the Minkowski dimension of a fractal string is the abscissa of
  convergence of its geometric zeta function \cite{La2}.

  To rephrase, we define the \emph{complex dimensions of \L} to be
  the set
  \begin{equation} \label{defn:complex-dimensions}
    \D = \{\gw \in \bC \suth \gz_\L \text{ is defined and has a pole at } \gw\}.
  \end{equation}

  One reason why these complex dimensions are important is the
  (explicit) tubular formula for fractal strings, a key result of
  \cite{La-vF1}. Namely, that under suitable conditions on the string
  \L, we have the following \emph{tube formula}:
  \begin{equation} \label{eqn:explicit-tubular-formula-for-strings}
    V(\ge) = \sum_{\gw \in \D} c_\gw \frac{(2\ge)^{1-\gw}}{\gw(1-\gw)} + R(\ge),
  \end{equation}
  where the sum is taken over the 
  complex dimensions \gw of \L, and the error term $R(\ge)$ is of
  lower order than the sum as $\ge \to 0^+$. (See \cite[Thm.~6.1, p. 144]{La-vF1}.) In the case when \L is a self-similar string
  i.e., when \del \L is a self-similar subset of \bR, one
  distinguishes two complementary cases (see [\citen{L},\citen{La3}]
  and \cite[\S2.3]{La-vF1} for further discussion of the
  lattice/nonlattice dichotomy):
  \begin{Comments}
    \item In the \emph{lattice case}, i.e., when the underlying
      scaling ratios are rationally dependent,
      the error term is shown to vanish identically and the complex
      dimensions lie periodically on vertical lines (including the
      line $\Re s = D$).

    \item In the \emph{nonlattice case}, the
      complex dimensions are quasiperiodically distributed and $s=D$ is
      the only complex dimension with real part $D$. Estimates for
      $R(\ge)$ are given in \cite[Thm.~6.20, p. 154]{La-vF1}
      and more precisely in [\citen{La-vF2}--3]. Also, \L is \Mink
      \measl if and only if it is nonlattice.
      See \cite[Chap. 2 and Chap. 6]{La-vF1} for details,
      including a discussion of quasiperiodicity.
  \end{Comments}

  These results pertain only to fractal subsets of \bR, although
  since this paper was completed and using a different approach, some
  preliminary work on the extension to the higher-dimensional case
  has been done in \cite{Pe} and [\citen{LaPe1}--2]. The foundational
  special case of `fractal sprays' \cite{LaPo2} is discussed in
  \cite[\S1.4]{La-vF1}.

\begin{figure}
    \scalebox{0.70}{\includegraphics{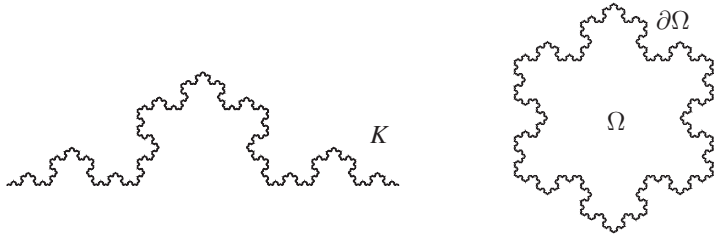}}
    \caption{The Koch curve $K$ and Koch snowflake domain \gW.}
    \label{fig:koch-snowflake} \centering
  \end{figure}

  It is the aim of the present paper to make some first steps
  in this direction. We compute $V(\ge)$ for a well-known
  (and well-studied) example, the Koch snowflake, with
  the hope that it may help in the development of a general
  higher-dimensional theory of complex dimensions.
  This curve provides an example of a \emph{lattice
  self-similar fractal} and a nowhere \diff plane curve.
  Further, the Koch snowflake can be viewed as the boundary
  \del\gW of a \bdd and simply connected open set $\gW \ci \bR^2$
  and that it is obtained by fitting together three congruent copies
  of the Koch curve $K$, as shown in Fig.~\ref{fig:koch-snowflake}.
  A general discussion of the Koch curve may be
  found in \cite[\S{}II.6]{Man} or \cite[Intro. and Chap. 9]{Fa}.

  The Koch curve is a self-similar fractal with dimension
  \(
    D := \log_3 4
  \) 
  (Hausdorff and Minkowski dimensions coincide
  for the Koch curve) and may be constructed by means of its self-similar
  structure (as in \cite[p. 15]{Ki} or \cite[Chap. 9]{Fa}) as follows:
  let \(\gr = \tfrac{1}{2} + \tfrac{1}{2\sqrt{3}}i,\)
  with $i=\sqrt{-1}$, and define two maps on \bC by
  \[f_1(z) := \gr \cj{z} \q \text{and} \q
    f_2(z) := (1-\gr)(\cj{z}-1)+1.\]
  Then the Koch curve is the self-similar set of $\bR^2$ with respect to
  $\{f_1,f_2\}$; i.e., the unique nonempty compact set $K \ci \bR^2$
  satisfying
  \(K = f_1(K) \cup f_2(K).\)

  In this paper, we prove the following new result:
  \begin{theorem}
    \label{thm:preview-of-final-result}
    The area of the inner \ge-\nbd of the Koch snowflake is
    given by the following tube formula:
    \begin{equation}
      \label{eqn:preview-of-final-result}
      V(\ge) = G_1(\ge) \ge^{2-D} + G_2(\ge) \ge^2,
    \end{equation}
    where $D=\log_34$ is the Minkowski dimension of \del\gW,
    $\per:= 2\gp/\log 3$ is the oscillatory period,
    and $G_1$ and $G_2$ are periodic \fns (of multiplicative
    period 3) which are discussed in full detail in
    Thm.~\ref{thm:final-result}. This formula may also be written
    \begin{equation}
      \label{eqn:preview-of-corollary}
      V(\ge) = \sum_{n \in \bZ} \gf_n \ge^{2-D-in\per} + \sum_{n \in \bZ} \gy_n \ge^{2-in\per},
    \end{equation}
    for suitable constants $\gf_n, \gy_n$ which depend only on
    $n$.  These constants are expressed in terms of the Fourier
    \coeffs $g_\ga$ of a multiplicative \fn which bears structural
    similarities to the classical Cantor--Lebesgue \fn described
    in more detail in \S\ref{subsec:formulating-h}.
  \end{theorem}

  While this formula is new, it should be noted that a previous approximation
  has been obtained in \cite[\S10.3]{La-vF1}; see
  (\ref{eqn:prev-V-of-eps-approx}) in Rem. \ref{rem:previous-formula-estimate}
  below. Our present formula, however, is exact. By reading off
  the powers of \ge appearing in \eqref{eqn:preview-of-corollary},
  we immediately obtain the following corollary:
  \begin{corollary}\label{cor:complex-dim-of-koch}
    The \possible complex dimensions of the Koch snowflake are
    \begin{equation}
      \label{eqn:complex-dim-of-koch}
      \D_{\del\gW} = \{D+in\per \suth n \in \bZ\}
        \cup \{in\per \suth n \in \bZ\}.
    \end{equation}
  \end{corollary}
  This is illustrated in Fig.~\ref{fig:complex-dimensions}.
  Also, for more precision regarding Cor. \ref{cor:complex-dim-of-koch},
  see Rem.~\ref{comment:cor-to-main-nondirect} below,
  as well as the discussion surrounding (\ref{eqn:poles by the picture}).

  \begin{remark}
    \label{rem:significance}
    \emph{
      The significance of the tube formula
      (\ref{eqn:preview-of-final-result}) is that it gives a
      detailed account of the oscillations that are intrinsic to
      the geometry of the Koch snowflake curve.
      More precisely, the \emph{real part}
      $D$ yields the order of the
      \emph{amplitude} of these oscillations (as a \fn of \ge)
      while the imaginary part $n\per = 2\gp n/\log3$
      $(n=0,1,2,\dots)$ gives their \emph{frequencies}. This is in
      agreement with the `philosophy' of the mathematical theory
      of the complex dimensions of fractal strings as developed in
      \cite{La-vF1}.
      Additionally, if one can show the existence of a complex
      dimension with real part $D$ and imaginary part $in\per, n \neq 0$,
      then Theorem~\ref{thm:preview-of-final-result} immediately
      implies that the Koch curve is not Minkowski measurable, as
      conjectured in \mbox{\cite[Conj.~2\&3, pp. 159,163--4]{La3}}.
      }
  \end{remark}

  The rest of this paper is dedicated to the proof of
  Thm.~\ref{thm:preview-of-final-result} (stated more precisely as
  Thm.~\ref{thm:final-result}).
  More specifically, in
  \S\ref{sec:estimating the area} we approximate the area
  $V(\ge)$ of the inner \ge-neighborhood.
  In \S\ref{sec:computing-the-error} we take into account the
  `error' resulting from this approximation. We study the form
  of this error in \S\ref{subsec:finding-the-area}, and the
  amount of it in \S\ref{subsec:counting-error-blocks}.
  In \S\ref{sec:computing-the-area} we combine
  \S\ref{sec:estimating the area} and \S\ref{sec:computing-the-error}
  to deduce the tube formula (\ref{eqn:preview-of-final-result})
  in the more precise form given in \S\ref{sec:results}.
  We also include in \S\ref{sec:results} some comments on the
  interpretation of Thm.~\ref{thm:final-result}.
  Finally, in \S\ref{subsec:formulating-h} we sketch the
  graph and briefly discuss some of the properties of the
  Cantor-like and multiplicatively periodic function $h(\ge)$, the
  Fourier coefficients of which occur explicitly in the expansion
  of $V(\ge)$ stated in Thm.~\ref{thm:final-result}.

\pgap

\emph{Acknowledgements.} The authors are grateful to Machiel van
Frankenhuysen for his comments on a preliminary version of this
paper. In particular, for indicating the current, more elegant and
concise presentation of the main result \eqref{eqn:final-result}.
We also wish to thank Victor Shapiro for a helpful discussion
concerning Fourier series, especially with regard to the
discussion surrounding \eqref{eqn:b times h}.

\section{Estimating the area}
  \label{sec:estimating the area}

  \begin{figure}[b]
    \scalebox{1.00}{\includegraphics{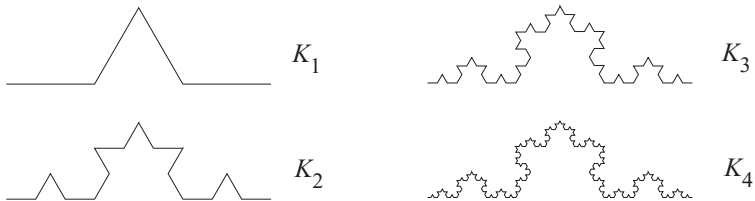}}
    \caption{The first four stages in the geometric construction of $K$.}
    \label{fig:koch-pre-curves} \centering
  \end{figure}

\begin{figure}[h]
    \scalebox{0.75}{\includegraphics{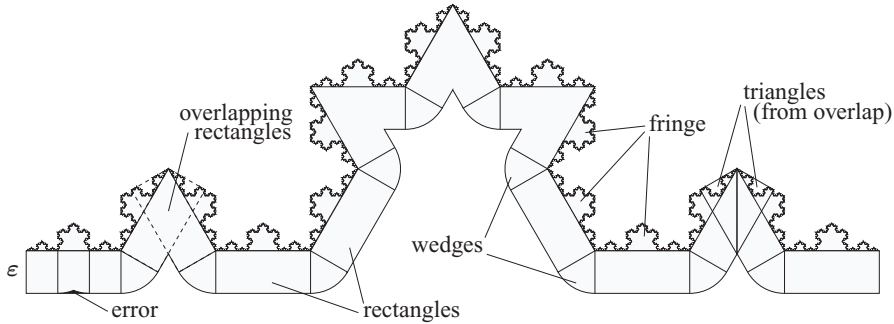}}
    \caption{An approximation to the inner \ge-\nbd of the Koch
    curve, with $\ge \in I_2$. The refinement level here is based on the graph $K_2$,
    the second stage in the geometric construction of the Koch
    curve (see Fig.~\ref{fig:koch-pre-curves}).}
    \label{fig:midneighborhood} \centering
  \end{figure}

  Consider an approximation to the inner \ge-\nbd of the Koch
  curve, as shown in Fig.~\ref{fig:midneighborhood}.
  Although we will eventually
  compute the neighborhood for the entire snowflake, we work with
  one third of it throughout the sequel (as depicted in the figure).

\begin{figure}[b]
    \scalebox{0.75}{\includegraphics{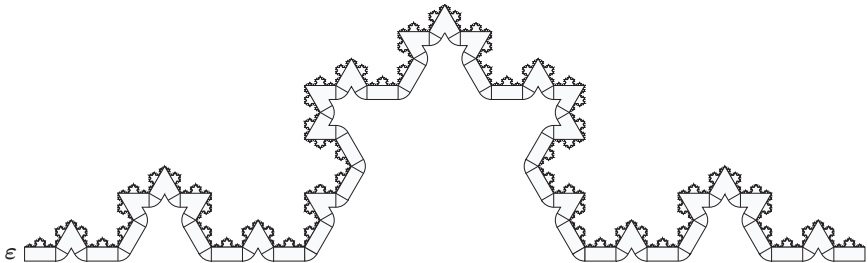}}
    \caption{A smaller \ge-\nbd of the Koch curve, for $\ge \in I_3$.
    This refinement level is based on the graph $K_3$, the third
    stage in the construction of the Koch curve.}
    \label{fig:neighborhood-refinements} \centering
  \end{figure}

  We will determine the area of the \ge-\nbd with functions that
  give the area of each kind of piece (rectangle, wedge, fringe,
  as shown in the figure) in terms of \ge, and functions that
  count the number of each of these pieces, in terms of \ge. As
  seen by comparing Fig.~\ref{fig:midneighborhood} to
  Fig.~\ref{fig:neighborhood-refinements}, the
  number of such pieces increases exponentially.

  \par We take the base of the Koch curve to have length 1, and
  carry out our approximation for different values of \ge,
  as $\ge \to 0^+$. In particular, let
  \begin{equation}\label{notation:I-of-n}
    I_n := (3^{-(n+1)}/\sqrt{3},3^{-n}/\sqrt{3}].
  \end{equation}

  Whenever $\ge = 3^{-n}/\sqrt{3}$, the approximation shifts to the
  next level of refinement. For example, Fig.~\ref{fig:midneighborhood}
  shows $\ge \in I_2$, and Fig.~\ref{fig:neighborhood-refinements}
  shows $\ge \in I_3$. Consequently, for $\ge \in I_0$, it suffices
  to consider an \ge-\nbd of the prefractal curve $K_0$ , and for $\ge \in I_1$, it suffices to
  consider an \ge-\nbd of the prefractal curve $K_1$, etc.
  Fig.~\ref{fig:koch-pre-curves} shows these prefractal approximations.

  In general, we will discuss a \nbd of $K_n$, and we define the function
  \begin{equation}\label{notation:n-of-epsilon}
    n = n(\ge)
      := \left[\log_3 \frac{1}{\ge\sqrt{3}}\right]
      = \left[x\right]
  \end{equation}
  to tell us for what $n$ we have $\ge \in I_n$. Here, the square brackets
  indicate the floor function (integer part) and
  \begin{equation}
    \label{notation:x-of-epsilon}
    x := -\log_3(\ge\sqrt{3}),
  \end{equation}
  a notation which will frequently prove convenient in the sequel.
  Further, we let
  \begin{equation}
    \label{notation:fractional-part-of-x}
    \{x\} := x-[x] \in [0,1)
  \end{equation}
  denote the fractional part of $x$.

  \par

  Observe that for $\ge \in I_n$, $n$ is fixed even as \ge is
  changing. To see how this is useful, consider that for all \ge in this
  interval, the number of rectangles (including those which
  overlap in the corners) is readily seen to be the fixed number
  \begin{equation}\label{notation:r-sub-n}
    r_n := 4^n.
  \end{equation}
  Also, each of these rectangles has area $\ge 3^{-n}$, where
  $n$ is fixed as \ge traverses $I_n$.
  Continuing in this constructive manner, we prove the following lemma.

  \begin{lemma}
    \label{lemma:4-components}
    For $\ge \in I_n$, there are
    \begin{Cond}
      \item $r_n = 4^n$ rectangles, each with area $\ge3^{-n}$,
      \item $w_n = \frac23(4^n-1)$ wedges, each with area $\frac{\gp\ge^2}{6}$,
      \item $u_n = \frac23(4^n+2)$ triangles, each with area $\frac{\ge^2\sqrt{3}}{2}$, and
      \item $4^n$ components of fringe, each with area $\frac{\sqrt3}{20}9^{-n}$.
    \end{Cond}
    \begin{proof}
      We have already established (i).
      \par For (ii), we exploit the self-similarity of the Koch curve $K$
      to obtain the recurrence relation
      \(w_n = 4w_{n-1}+2\),
      which we solve to find the number of wedges
      \begin{equation}\label{notation:w-sub-n}
        w_n := \sum_{j=0}^{n-1}{2\cdot 4^j}
          = \tfrac{2}{3}\left(4^n - 1\right).
      \end{equation}
      The area of each wedge is clearly
      \(\gp\ge^2/6,\)
      as the angle is always fixed at $\gp/3$.
      \par (iii)   To prevent double-counting, we will need to keep track of the
      number of rectangles that overlap in the acute angles so that we
      may subtract the appropriate number of triangles
      \begin{equation}\label{notation:u-sub-n}
        u_n := 4^n - \sum_{j=1}^{n-1} 4^j
          = \tfrac{2}{3}\left(4^n + 2\right).
      \end{equation}
      Each of these triangles has area $\ge^2 \sqrt{3}/2$.
      \par (iv) To measure the area of the fringe, note that the area
      under the entire Koch curve is given by $\sqrt{3}/20$,
      so the fringe of $K_n$ will be this number scaled by $\left(3^{-n}\right)^2$.
      There are $4^n$ components, one atop each rectangle (see Fig.~\ref{fig:midneighborhood}).
    \end{proof}
  \end{lemma}
  \par
  Lemma~\ref{lemma:4-components} gives a preliminary area
  formula for the \ge-\nbd. Here, `preliminary' indicates the
  absence of the `error estimate' developed in
  \S\ref{sec:computing-the-error}.
  \begin{lemma}
    \label{lemma:approximate area}
    The \ge-\nbd of the Koch curve has approximate area
    \begin{equation}\label{fn:preliminary-area} 
      \preV(\ge) = \ge^{2-D} 4^{-\{x\}}
        \left(\tfrac{3\sqrt{3}}{40}9^{\{x\}} + \tfrac{\sqrt{3}}{2} 3^{\{x\}}
        + \tfrac16\left(\tfrac{\gp}{3} - \sqrt{3}\right)\right)
        - \tfrac{\ge^2}3 \left(\tfrac{\gp}{3} + 2\sqrt{3}\right).
    \end{equation}
  \end{lemma}
    This formula is \emph{approximate} in the sense that it measures a region
    slightly larger than the actual \ge-\nbd. This discrepancy is
    accounted for and analyzed in detail in \S3.
    \begin{proof}[of Lemma~\ref{lemma:approximate area}]
      Using \eqref{notation:n-of-epsilon} and \eqref{notation:x-of-epsilon},
      we obtain:
      \begin{equation}
        \label{eqn:identities-with-const-to-the-x}
        4^x = \tfrac{1}{2}\ge^{-D}, \hstr
        9^{-x}=3\ge^2, \hstr
        \left(\tfrac{4}{3}\right)^x = \tfrac{\sqrt{3}}{2}\ge^{1-D}, \hstr
        \left(\tfrac{4}{9}\right)^x = \tfrac{3}{2}\ge^{2-D}.
      \end{equation}
      Now using $n=[x]=x-\{x\}$, we compute the contributions of
      the rectangles, wedges, triangles, and fringe, respectively, as
      \begin{align*}
        \preV_r(\ge) &= \ge\left(\tfrac{4}{3}\right)^n = \ge^{2-D}\tfrac{\sqrt{3}}{2} \cdot 4^{-\{x\}} 3^{\{x\}},\\
        \preV_w(\ge) &= \tfrac{\gp\ge^2}{9}(4^n - 1) = \ge^{2-D}\tfrac{\gp}{18} \cdot 4^{-\{x\}} - \ge^2\tfrac{\gp}{9},\\
        \preV_u(\ge) &= \tfrac{\ge^2\sqrt{3}}{3}\left(4^n+2\right) = \ge^{2-D}\tfrac{\sqrt3}6 \cdot 4^{-\{x\}}+\ge^2\tfrac{2\sqrt3}3, \text{ and}\\
        \preV_{\text{f}}(\ge)
         &= \left(\tfrac{4}{9}\right)^n \left(\tfrac{\sqrt{3}}{20}\right)
          = \ge^{2-D}\tfrac{3\sqrt{3}}{40} \cdot 4^{-\{x\}} 9^{\{x\}}.
      \end{align*}
      Putting all this together,
      $\preV = \preV_r + \preV_w - \preV_u + \preV_{\text{f}}$ gives the result.
    \end{proof}

  \par

  \begin{remark}
    \label{rem:previous-formula-estimate}
    \emph{
    It is pleasing to find that this is in agreement with earlier
    predictions of what $\preV(\ge)$ should look like. In particular,
    \cite[p. 209]{La-vF1} gives the estimate
    \begin{equation} \label{eqn:prev-V-of-eps-approx}
      V(\ge) \approx \ge^{2-D} \frac{\sqrt{3}}{4} 4^{-\{x\}} \left(\frac{3}{5}9^{\{x\}} + 6 \cdot 3^{\{x\}} - 1\right),
    \end{equation}
    which differs only from our formula for $\preV(\ge)$ in  (\ref{fn:preliminary-area}) by some constants and the
    final term of order $\ge^2$.
    }
  \end{remark}

  \par In  (\ref{eqn:E(eps) simplified}) we will
  require the Fourier series of the periodic \fn
  $\ge^{-(2-D)}\preV(\ge)$ as
  it is given by  (\ref{fn:preliminary-area}),
  so we recall the formula
  \begin{equation} \label{eqn:b-to-frac-u-fourier-series-formula}
    a^{-\{x\}}
    = \frac{a-1}{a} \sum_{n \in \bZ} \frac{e^{2\gp inx}}{\log a + 2\gp in}.
  \end{equation}
  This formula is valid for $a>0, a \neq 1$ and has been used repeatedly
  in \cite{La-vF1}. Note that it follows from Dirichlet's Theorem and
  thus holds in the sense of Fourier series. In particular,
  the series in \eqref{eqn:b-to-frac-u-fourier-series-formula} converges
  pointwise; this is also true in \eqref{eqn:fourier-expn-of-a-to-the-frac-x}
  and \eqref{eqn:preliminary-fourier-series} below.

  We will make frequent use of the following identity in the sequel:
  \begin{equation}
    \label{eqn:simplified-fourier-term}
    e^{2\gp inx}
    = \left(\ge\sqrt3\right)^{-in\per}
    = (-1)^n \ge^{-in\per},
    \q \text{ for } n \in \bZ,
  \end{equation}
  where $\per = 2\gp/\log3$ is the oscillatory period as in
  Thm.~\ref{thm:preview-of-final-result}.
  With \makebox{$x=-\log_3(\ge\sqrt3)$}, we can rewrite the
  Fourier expansion of $a^{-\{x\}}$ given in
  \eqref{eqn:b-to-frac-u-fourier-series-formula} as
  \begin{equation}
    \label{eqn:fourier-expn-of-a-to-the-frac-x}
    a^{-\{x\}} = \frac{a-1}{a \log3} \sum_{n \in \bZ} \frac{(-1)^n\ge^{-in\per}}{\log_3 a + in\per}.
  \end{equation}

  Recall that $\{x\}=x-[x]$ denotes the fractional part
  of $x$. With $D=\log_34$, we use
  \eqref{eqn:fourier-expn-of-a-to-the-frac-x} to express
  \eqref{fn:preliminary-area}
  as a pointwise convergent Fourier series in \ge:
  \begin{align}
    \preV(\ge)
    &= \tfrac1{3\log3} \sum_{n \in \bZ}
      \left(\tfrac{-3^{5/2}}{2^5(D - 2 + in\per)}
      + \tfrac{3^{3/2}}{2^3(D - 1 + in\per)}
      + \tfrac{\gp - 3^{3/2}}{2^3(D + in\per)} \right)
      (-1)^n \ge^{2-D-in\per}
      \notag \\ & \hstr[50]
      - \tfrac13\left(\tfrac{\gp}{3} + 2\sqrt{3}\right) \ge^2.
      \label{eqn:preliminary-fourier-series}
  \end{align}

\section{Computing the error}
  \label{sec:computing-the-error}

\begin{figure}
    \scalebox{0.80}{\includegraphics{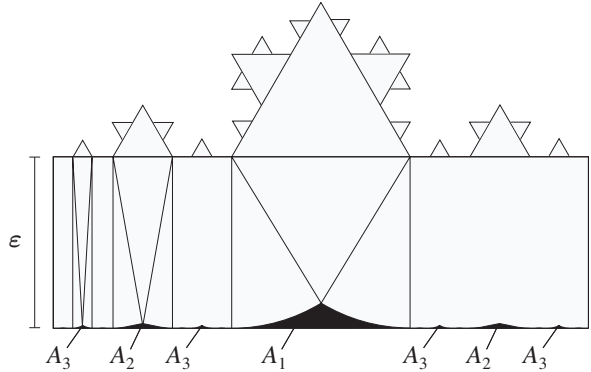}}
    \caption{An error block for $\ge \in I_n$. The central third of
    the block contains one large isosceles triangle, two wedges,
    and the trianglet $A_1$.}
    \label{fig:error-block} \centering
  \end{figure}

  Now we must account for all the little `trianglets', the small
  regions shaped like a crest of water on an ocean wave. These regions
  were included in our original calculation, but now must be
  subtracted. This error appeared in each of the rectangles counted
  earlier, and so we refer to all the error from one of these
  rectangles as an `error block'. Fig.~\ref{fig:error-block}
  shows how this error is incurred and how it inherits a Cantoresque
  structure from the Koch curve.
  Actually, we will see in \S\ref{subsec:counting-error-blocks}
  that this `error' will have several terms, some of which are of the same
  order as the leading term in $\preV(\ge)$, which is proportional
  to $\ge^{2-D}$ by \eqref{eqn:preliminary-fourier-series}.
  Hence, caution should be exercised when carrying
  out such computations and one should not be too quick to set
  aside terms that appear negligible.

\subsection{Finding the area of an `error block'}
  \label{subsec:finding-the-area}
\begin{figure}[b]
    \scalebox{0.70}{\includegraphics{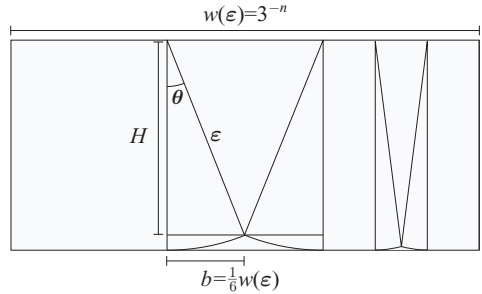}}
    \caption{Finding the height of the central triangle.}
    \label{fig:finding-the-height} \centering
  \end{figure}

  In calculating the error, we begin by finding the area of one of
  these error blocks. Later, we will count how many of these error
  blocks there are, as a function of \ge. Note that $n$ is fixed
  throughout \S\ref{subsec:finding-the-area}, but \ge varies within $I_n$.
  We define the function
  \begin{equation}\label{fn:w-of-epsilon}
    w = w(\ge) := 3^{-n}
      = 3^{-[x]}.
  \end{equation}
  This function $w(\ge)$ gives the width of one of the rectangles,
  as a function of \ge (see Fig.~\ref{fig:finding-the-height}).
  Note that $w(\ge)$ is constant as \ge traverses $I_n$, as is $n=n(\ge)$.
  From Fig.~\ref{fig:finding-the-height}, one can work out that
  the area of both wedges adjacent to $A_k$ is
  \begin{equation*} 
    \ge^2\asin\left(\tfrac{w}{2\cdot3^k\ge}\right),
  \end{equation*}
  and that the area of the triangle above $A_k$ is
  \begin{equation*} 
    \ge \tfrac{w}{2\cdot3^k} \sqrt{1-\left(\tfrac{w}{2\cdot3^k\ge}\right)^2}.
  \end{equation*}

  \par Then for $k=1,2,\dots,$ the area of the trianglet $A_k$
  is given by
  \begin{equation} \label{eqn:kth-trianglet-area}
    A_k(\ge) = \ge\frac{w(\ge)}{3^k}
        - \ge^2\asin\left(\frac{w(\ge)}{2\cdot 3^k \ge}\right)
        - \ge \frac{w(\ge)}{2 \cdot 3^k} \sqrt{1-\left(\frac{w(\ge)}{2\cdot 3^k \ge}\right)^2},
  \end{equation}
  and appears with multiplicity $2^{k-1}$, as in Fig.~\ref{fig:error-block}.
  We use \eqref{fn:w-of-epsilon} and \eqref{notation:x-of-epsilon} to write
  \(w(\ge) = 3^{-x}3^{\{x\}}
    = \ge\sqrt{3}(\tfrac{1}{3})^{-\{x\}},\)
  and define
  \begin{equation}
    \threexk := \tfrac{w}{3^k\ge} = 3^{\{x\}-k+1/2}.
    \label{notation:threexk}
  \end{equation}
  Hence the entire contribution of one error block may be written as
  \begin{align}
    \label{eqn:error-block-area}
    B(\ge):&= \sum_{k=1}^\iy 2^{k-1}
      \left(\threexk
        - \asin\left(\tfrac\threexk2\right)
        - \tfrac\threexk2 \sqrt{1-\left(\tfrac\threexk2\right)^2}
        \right) \ge^2.
  \end{align}
  Recall the power series expansions
  \[\asin u = \sum_{m=0}^\iy {\tfrac{(2m)! \, u^{2m+1}}{2^{2m}(m!)^2(2m+1)}}
    \q \text{ and } \q
    \sqrt{1-u^2} = 1-\sum_{m=0}^\iy {\tfrac{(2m)! \, u^{2m+2}}{2^{2m+1}m!(m+1)!}},\]
  which are valid for $|u|<1$.
  We use these formulae with $u=\frac w{2\cdot 3^k \ge}$, so
  convergence is guaranteed by
  \[0 \leq \tfrac{w}{2\cdot3^k\ge}
    = \tfrac{3^{\{x\}}\sqrt3}{2\cdot3^k}
    \leq \tfrac{\sqrt3}{2} < 1,\]
  and the fact that the series in \eqref{eqn:error-block-area}
  starts with $k=1$. Then
  \eqref{eqn:error-block-area} becomes
  \begin{align}
    \label{eqn:error-block-area-simplified}
    B(\ge)
      &= \sum_{k=1}^\iy 2^{k-1}
        \left[\tfrac\threexk2
          + \sum_{m=0}^\iy \tfrac{(2m)! \, (\threexk)^{2m+3}}{2^{4m+4}m!(m+1)!}
          - \sum_{m=0}^\iy \tfrac{(2m)! \, (\threexk)^{2m+1}}{2^{4m+1}(m!)^2(2m+1)}
        \right]\ge^2 \notag \\
      &= \sum_{k=1}^\iy 2^{k-1}
        \left[
          \sum_{m=1}^\iy \tfrac{(2m-2)! \, (\threexk)^{2m+1}}{2^{4m}(m-1)!m!}
          - \sum_{m=1}^\iy \tfrac{(2m)! \, (\threexk)^{2m+1}}{2^{4m+1}(m!)^2(2m+1)}
        \right]\ge^2 \notag \\
      &= \sum_{m=1}^\iy \sum_{k=1}^\iy
          \tfrac{2^{k-1}}{(3^{2m+1})^k}
          \tfrac{(2m-2)! \, (3^{\{x\}+1/2})^{2m+1}}{2^{4m}(m-1)!m!}
          \left(1-\tfrac{(2m-1)2m}{2m(2m+1)}\right)
        \ge^2 \notag \\
      &= \sum_{m=1}^\iy \tfrac1{3^{2m+1}}\left(\tfrac1{(3^{2m+1}-2)/3^{2m+1}}\right)
          \tfrac{(2m-2)! \, (\sqrt3)^{2m+1}}{2^{4m-1}(m-1)!m!(2m+1)}
          \left(\tfrac1{3^{2m+1}}\right)^{-\{x\}}
        \ge^2 \notag \\
      &= \sum_{m=1}^\iy
          \tfrac{(2m-2)!}
            {2^{4m-1} (m-1)! m! (2m+1) (3^{2m+1}-2) }
          \left(\tfrac1{3^{2m+1}}\right)^{-\{x\}}
        \ge^2.
  \end{align}
  The interchange of sums is validated by checking absolute
  convergence of the final series via the ratio test,
  and then applying Fubini's Theorem to retrace our steps.

\subsection{Counting the error blocks}
  \label{subsec:counting-error-blocks}
\begin{figure}[b]
    \scalebox{0.85}{\includegraphics{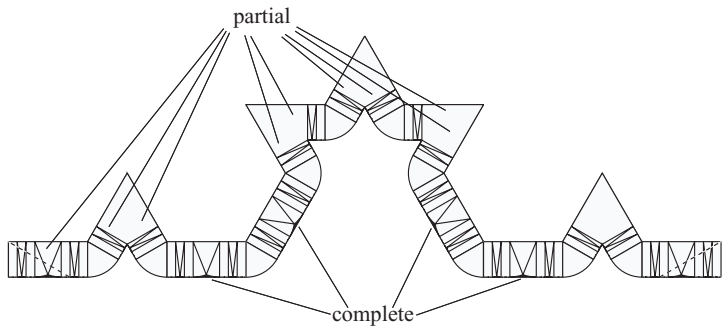}}
    \caption{Error block formation. The ends are counted as partial because
    three of these pieces will be added together to make the entire snowflake.}
    \label{fig:forming-blocks} 
  \end{figure}

  Some blocks are present in their entirety as \ge traverses an
  interval $I_n$, while others are in the process of forming:
  two in each of the peaks and one at each end
  (see Fig.~\ref{fig:forming-blocks}).
  Using the same notation as previously in Lemma \ref{lemma:4-components},
  we count the complete and partial error blocks with
  \begin{align*}
    c_n &= r_n - u_n
      =\tfrac{1}{3}\left(4^n-4\right),
      \q \text{and} \q
    p_n = u_n
      = \tfrac{2}{3}\left(4^n+2\right).
  \end{align*}

  By means of (\ref{notation:n-of-epsilon})--(\ref{notation:fractional-part-of-x}),
  we convert $c_n$ and $p_n$ into \fns of the \cn variable \ge, where $\ge >0$:
  \begin{align}
    \label{eqn:c-and-p-as-contin-fns}
    c(\ge)
      &= \tfrac13 \left(\tfrac{\ge^{-D}}{2}4^{-\{x\}}-4\right)
       \q \text{ and } \q
    p(\ge)
       = \tfrac23 \left(\tfrac{\ge^{-D}}{2}4^{-\{x\}}+2\right).
  \end{align}

  With $B(\ge)$ given by \eqref{eqn:error-block-area-simplified},
  the total error is thus
  \begin{equation}
    \label{notation:E-of-epsilon}
    E(\ge) = B(\ge)\left[c(\ge) + p(\ge)h(\ge)\right].
  \end{equation}

\pgap

  \begin{remark}
    \label{rem:introduction of h}
    \emph{
    The \fn $h(\ge)$ in \eqref{notation:E-of-epsilon} is some
    periodic function that oscillates multiplicatively in a region bounded between
    0 and 1, indicating what portion of the partial error block has
    formed; see Fig.~\ref{fig:error-block} and Fig.~\ref{fig:forming-blocks}.
    We do not know $h(\ge)$ explicitly, but we do know by the self-similarity
    of $K$ that it has multiplicative period 3; i.e., $h(\ge) =
    h(\tfrac\ge3)$. Using \eqref{eqn:simplified-fourier-term},
    the Fourier expansion
    \begin{equation}
      \label{eqn:prelim-fourier-expn-of-h}
      h(\ge) = \sum_{\ga \in \bZ} g_\ga (-1)^\ga\ge^{-i\ga\per}
        = \sum_{\ga \in \bZ} g_\ga e^{2\gp i \ga x} = g(x)
    \end{equation}
    shows that we may also consider $h(\ge)$ as an \emph{additively}
    periodic \fn of the variable $x=-\log_3(\ge\sqrt3)$, with
    additive period 1.
    We refer the interested reader to \S\ref{subsec:formulating-h} below for a further discussion of $h(\ge)$,
    including a sketch of its graph, justification of the convergence of
    \eqref{eqn:prelim-fourier-expn-of-h}, and a brief
    discussion of some of its properties.
    }
  \end{remark}

  We now return to the computation; substituting
  \eqref{eqn:error-block-area-simplified} and
  \eqref{eqn:c-and-p-as-contin-fns} into \eqref{notation:E-of-epsilon}
  gives
  \begin{align}
    E(\ge)
      &= B(\ge)\left[\tfrac{\ge^{-D}}3 4^{-\{x\}}\left(\tfrac12 + h(\ge)\right) + \tfrac43 \left(h(\ge)-1\right) \right]
        \notag \\
      &= \tfrac13 \sum_{m=1}^\iy
        \tfrac{(2m-2)!  (h(\ge) + 1/2)}
         {2^{4m-1} (m-1)! m! (2m+1) (3^{2m+1}-2)}
        \left(\tfrac4{3^{2m+1}}\right)^{-\{x\}} \ge^{2-D} \notag \\
      &\hstr[13]
       +\tfrac13
        \sum_{m=1}^\iy
        \tfrac{(2m-2)! (h(\ge) - 1)}
         {2^{4m-3} (m-1)! m! (2m+1) (3^{2m+1}-2)}
        \left(\tfrac1{3^{2m+1}}\right)^{-\{x\}} \ge^{2} \notag \\
      &= \tfrac1{3\log3}
        \sum_{m=1}^\iy \sum_{n \in \bZ}
        \tfrac{(2m-2)! (4-3^{2m+1}) (-1)^n (h(\ge) + 1/2)}
         {2^{4m+1} (m-1)! m! (2m+1) (3^{2m+1}-2) (D-2m-1+in\per)}
        \ge^{2-D-in\per} \notag \\
      &\hstr[13]
       + \tfrac1{3\log3}
        \sum_{m=1}^\iy \sum_{n \in \bZ}
        \tfrac{(2m-2)! (1-3^{2m+1}) (-1)^n (h(\ge) - 1)}
         {2^{4m-3} (m-1)! m! (2m+1) (3^{2m+1}-2) (-2m-1+in\per)}
        \ge^{2-in\per} \notag \\
      &= \tfrac1{3\log3}
        \sum_{n \in \bZ} (h(\ge) + 1/2) (-b_n)
        (-1)^n \ge^{2-D-in\per} \notag \\
        &\hstr[13]
       + \tfrac1{3\log3}
        \sum_{n \in \bZ} (h(\ge) - 1) (-\gt_n)
        (-1)^n \ge^{2-in\per},
      \label{eqn:E(eps) simplified}
  \end{align}
  where we have written the constants $b_n$ and $\gt_n$
  in the shorthand notation as follows:
  \begin{align}
    \label{notation:b_n}
    b_n &:= \sum_{m=1}^\iy
        \tfrac{(2m-2)! (3^{2m+1}-4) }
         {2^{4m+1} (m-1)! m! (2m+1) (3^{2m+1}-2) (D-2m-1+in\per)}, \\
    \label{notation:tau_n}
    \text{ and }
    \gt_n &:= \sum_{m=1}^\iy
        \tfrac{(2m-2)! (3^{2m+1}-1) }
         {2^{4m-3} (m-1)! m! (2m+1) (3^{2m+1}-2) (-2m-1+in\per)}.
  \end{align}
  In the third equality of \eqref{eqn:E(eps) simplified}, we have
  applied \eqref{eqn:fourier-expn-of-a-to-the-frac-x} to
  $a=4/3^{2m+1}$ and to \makebox{$a=1/3^{2m+1}$}, respectively.
  By the ratio test, the complex numbers $b_n$ and $\gt_n$ given by
  \eqref{notation:b_n} and \eqref{notation:tau_n} are \wdef. This
  fact, combined with Fubini's Theorem, enables us to justify the
  interchange of sums in the last equality of
  \eqref{eqn:E(eps) simplified}.

\section{Computing the area}
  \label{sec:computing-the-area}

  Now that we have estimate \eqref{eqn:preliminary-fourier-series}
  for the area of the \nbd of the Koch curve $\preV(\ge)$,
  and formula \eqref{eqn:E(eps) simplified} for the `error' $E(\ge)$,
  we can find the exact area of the inner \nbd of the full Koch
  snowflake as follows:
  \begin{align}
    V(\ge)
    &= 3\left(\preV(\ge) - E(\ge)\right) \notag \\
    &= \tfrac1{\log3} \sum_{n \in \bZ}
      \left(\tfrac{-3^{5/2}}{2^5(D-2+in\per)}
            + \tfrac{3^{3/2}}{2^3(D-1+in\per)}
            + \tfrac{\gp - 3^{3/2}}{2^3(D+in\per)}\right)
      (-1)^n \ge^{2-D-in\per}
        \notag \\
      &\hstr[5]
       +\tfrac1{\log3}
        \sum_{n \in \bZ} (h(\ge) + 1/2) (-1)^n b_n
        \ge^{2-D-in\per}
        \notag \\
      &\hstr[5]
       - \left(\tfrac\gp3 + 2\sqrt3\right) \ge^2
       + \tfrac1{\log3}
        \sum_{n \in \bZ} (h(\ge) - 1) (-1)^n \gt_n
        \ge^{2-in\per}
        \notag \\
    &= \tfrac1{\log3} \sum_{n \in \bZ}
      \left(\tfrac{-3^{5/2}}{2^5(D-2+in\per)}
            + \tfrac{3^{3/2}}{2^3(D-1+in\per)}
            + \tfrac{\gp - 3^{3/2}}{2^3(D+in\per)}
            \right. \notag \\ &\hstr[36] \left. \vstr[2.2]
            + \tfrac{b_n}2
            + h(\ge)b_n\right)
      (-1)^n \ge^{2-D-in\per}
        \notag \\
      &\hstr[5]
       + \tfrac1{\log3}
        \sum_{n \in \bZ} \left(-\gt_n - \log3\left(\tfrac\gp3 + 2\sqrt3\right)\gd_0^n
          + h(\ge)\gt_n\right) (-1)^n
        \ge^{2-in\per},
    \label{eqn:unsimplified full area}
  \end{align}
  where $\gd_0^n$ is the Kronecker delta.
  Therefore, we have
  \begin{equation}
    \label{eqn:final-result}
    V(\ge) = G_1(\ge) \ge^{2-D} + G_2(\ge) \ge^2,
  \end{equation}
  where the periodic \fns $G_1$ and $G_2$ are given by
  \begin{align}
    \label{eqn:first periodic fn, introduced}
    G_1(\ge) :&= \frac1{\log3} \sum_{n \in \bZ}
      \left(a_n + b_n h(\ge) \right)
      (-1)^n \ge^{-in\per}
    \\
    \label{eqn:second periodic fn, introduced}
    \text{ and \;}
    G_2(\ge) :&= \frac1{\log3} \sum_{n \in \bZ}
      \left(\gs_n + \gt_n h(\ge) \right)
      (-1)^n \ge^{-in\per}.
  \end{align}
  Here we have used \eqref{notation:b_n} and \eqref{notation:tau_n},
  and introduced the notation $a_n$ and $\gs_n$ (stated explicitly
  in \eqref{eqn:mean-coeffs}).
  We wish to rearrange these series so as to collect all factors of \ge.
  First, we can split the sum in \eqref{eqn:first periodic fn, introduced} as
  \begin{align}
    \label{eqn:G1 after split}
    G_1(\ge)
    &= \frac1{\log3} \sum_{n \in \bZ} a_n (-1)^n \ge^{-in\per}
      + \frac{h(\ge)}{\log3} \sum_{n \in \bZ} b_n (-1)^n \ge^{-in\per}
  \end{align}
  because
  \(a(\ge) := \sum_{n \in \bZ} a_n (-1)^n \ge^{-in\per}\)
  and
  \(b(\ge) := \sum_{n \in \bZ} b_n (-1)^n \ge^{-in\per}\)
  are each convergent: $a(\ge)$ converges for the same reason as
  \eqref{eqn:preliminary-fourier-series}, and one can show that
  \begin{align}
    \label{eqn:b(eps) as sum}
    b(\ge) = \sum_{m=0}^\iy
      \tfrac{(4\log3) (2m-2)!}
      {2^{4m+1} 3^{m-1/2} (m-1)! m! (2m+1) (3^{2m+1}-2)}
      \left(\tfrac{4}{3^{2m+1}}\right)^{-\{x\}}
  \end{align}
  converges to a well-defined distribution induced by a locally
  integrable \fn; one proves directly that $|b_n| \leq c/|n|$ by writing
  \eqref{notation:b_n} as
  \begin{align}
    \label{eqn:bn as conv sum}
    b_n 
        &= \sum_{m=1}^\iy \tfrac{\gb_m}{D-2m-1+in\per},
        \q \text{ with } \q
        \sum_{m=1}^\iy \gb_m \, < \, \iy.
  \end{align}
  Then the rearrangement leading to \eqref{eqn:b(eps) as sum} is justified via
  the ``descent method'' with $q=2$, as described in Rem.~\ref{rem:convergence}
  below.
  Note that the right-hand side of \eqref{eqn:b(eps) as sum}
  converges by the ratio test and is thus defined pointwise on
  $\bR \less \bZ$.
  Therefore one also sees that $b(\ge)$ and $h(\ge)$ are periodic \fns.
  Considered as functions of the variable $x = \log_3(1/\ge\sqrt3)$,
  both have period 1 with $b$ \cn for $0 \leq x < 1$ and $h$
  \cn for $0 < x \leq 1$. Further, each is monotonic on its period interval,
  and possesses a bounded jump discontinuity only at the endpoint.
  This may be seen for $b(\ge)$ from \eqref{eqn:b(eps) as sum} and
  for $h(\ge)$ from \S6.

  Although we have used \distl arguments to obtain \eqref{eqn:b(eps) as sum},
  the right-hand side of \eqref{eqn:b(eps) as sum} is locally
  integrable and has a representation as a piecewise \cn
  \fn. Thus the \distl equality in \eqref{eqn:b(eps) as sum}
  actually holds pointwise in the sense of Fourier series and
  all our results are still valid pointwise.
  Recall that the Dirichlet--Jordan Theorem \cite[Thm. II.8.1]{Zy}
  states that if $f$ is periodic and (locally) of bounded variation,
  its Fourier series converges pointwise to
  $\left(f(x-)+\vstr[2]f(x+)\right)/2$.
  As described above, $b(\ge)$ and $h(\ge)$ are each of bounded
  variation and therefore by {\cite[Thm. II.4.12]{Zy}} we have
  \(
    b_n = O(1/n) \text{ and } g_n = O(1/n)
    \text{ as } n \to \pm \iy.
  \) 
  Finally, \cite[Thm. IX.4.11]{Zy} may be applied to yield the
  pointwise equality
  \begin{align}
    b(\ge)h(\ge)
    = \sum_{n \in \bZ} \sum_{\ga \in \bZ}
          b_\ga g_{n-\ga} (-1)^{n} \ge^{-in\per}.
    \label{eqn:b times h}
  \end{align}
  This theorem applies \bc \eqref{eqn:b(eps) as sum} shows that
  $b(\ge)$ is \bdd away from 0.

  Now that all the \ge's are combined, we substitute \eqref{eqn:b times h}
  back into \eqref{eqn:G1 after split} and rewrite
  $G_1(\ge) = \left(a(\ge) + \vstr[2]b(\ge)h(\ge)\right)/\log3$ as
  in \eqref{eqn:first periodic fn}:
  \begin{align}
    \label{eqn:G1 in final form}
    G_1(\ge)
    &= \frac1{\log3} \sum_{n \in \bZ}
            \left(a_n + \sum_{\ga \in \bZ} b_\ga g_{n-\ga} \right)
            (-1)^n \ge^{-in\per}.
  \end{align}

Manipulating $G_2(\ge)$ similarly, we are able to rewrite
\eqref{eqn:second periodic fn, introduced} in its final form
\eqref{eqn:second periodic fn}, and thereby complete the proof of
Thm.~\ref{thm:final-result}.

\begin{remark}
  \label{rem:convergence}
  \emph{
  The convergence of the Fourier series associated to a
  periodic distribution is proved via the descent method by
  integrating both sides $q$ times (for sufficiently large $q$)
  so that one has pointwise convergence.
  After enough integrations, the distribution will be a smooth \fn,
  the series involved will converge absolutely, and the
  Weierstrass theorem can be applied pointwise.
  At this point, rearrangements or interchanges of series are justified
  pointwise, and we obtain a pointwise formula for the \nth[q] antiderivative
  of the desired function. Then one takes the distributional
  derivative $q$ times to obtain the desired formula.
  See \cite[Rem.~4.14]{La-vF1}. 
  How large the positive integer $q$ needs to be depends on 
  the order of
  polynomial growth of the Fourier coefficients.
  Recall that the Fourier series of a periodic \dist \cvs \distly if and
  only if the Fourier \coeffs are of slow growth, i.e., do not grow
  faster than polynomially. Moreover, from the point of view of
  \dists, there is no distinction to be made \btwn \cvn \trig series
  and Fourier series. 
  See \cite[\S{}VII,I]{Sch1}.
  }
\end{remark}

\section{Main results}
  \label{sec:results}

  We can now state our main result in the following more precise
  form of Thm.~\ref{thm:preview-of-final-result}:
  \begin{theorem}
    \label{thm:final-result}
    The area of the inner \ge-\nbd
    of the Koch snowflake is given pointwise by the following tube formula:
      \begin{equation}
        \label{eqn:final-result}
        V(\ge) = G_1(\ge) \ge^{2-D} + G_2(\ge) \ge^2,
      \end{equation}
      where $G_1$ and $G_2$ are periodic \fns of multiplicative
      period 3, given by
      \begin{subequations}
        \label{eqngrp:the periodic fns}
        \begin{align}
          \label{eqn:first periodic fn}
          G_1(\ge) :&= \frac1{\log3} \sum_{n \in \bZ}
            \left(a_n + \sum_{\ga \in \bZ} b_\ga g_{n-\ga} \right)
            (-1)^n \ge^{-in\per}
          \\
          \label{eqn:second periodic fn}
          \text{ and \;} G_2(\ge) :&= \frac1{\log3} \sum_{n \in \bZ}
            \left(\gs_n + \sum_{\ga \in \bZ} \gt_\ga g_{n-\ga} \right)
            (-1)^n \ge^{-in\per},
        \end{align}
      \end{subequations}
    where
    $a_n, b_n, \gs_n,$ and $\gt_n$ are
    the complex numbers given by
      \begin{align}
        a_n &= -\frac{3^{5/2}}{2^5(D - 2 + in\per)}
        +\frac{3^{3/2}}{2^3(D - 1 + in\per)}
        +\frac{\gp - 3^{3/2}}{2^3(D + in\per)}
        +\frac12 b_n,
        \notag \\
        b_n &= \sum_{m=1}^\iy
         \frac{ (2m)! \; (3^{2m+1}-4)}
           {4^{2m+1} (m!)^2 (4m^2-1) (3^{2m+1}-2) (D - 2m - 1 + in\per)},
        \notag \\
        \label{eqn:mean-coeffs}
        \gs_n &= - \log3\left(\frac{\gp}{3} + 2\sqrt{3}\right)\gd_0^n
          - \gt_n, \text{ and } \\
        \gt_n &= \sum_{m=1}^\iy
         \frac{(2m)! \; (3^{2m+1}-1)}
           {4^{2m-1} (m!)^2 (4m^2-1) (3^{2m+1}-2) (-2m - 1 + in\per)},
        \notag
       \end{align}
    where $\gd_0^{\,0}=1$ and $\gd_0^n=0$ for $n \neq 0$ is the
    Kronecker delta.
  \end{theorem}

    In Thm.~\ref{thm:final-result}, $D = \log_34$ is the \Mink
    dimension of the Koch snowflake
    \del\gW and $\per=2\gp/\log3$ is its oscillatory period,
    following the terminology of \makebox{\cite{La-vF1}}. The numbers
    $g_\ga$ appearing in \eqref{eqngrp:the periodic fns} are the Fourier \coeffs of the periodic \fn $h(\ge)$, a
    suitable nonlinear analogue of the Cantor--Lebesgue function,
    defined in Rem. \ref{rem:introduction of h} and further discussed in
    \S\ref{subsec:formulating-h} below.

\begin{figure}
    \scalebox{0.80}{\includegraphics{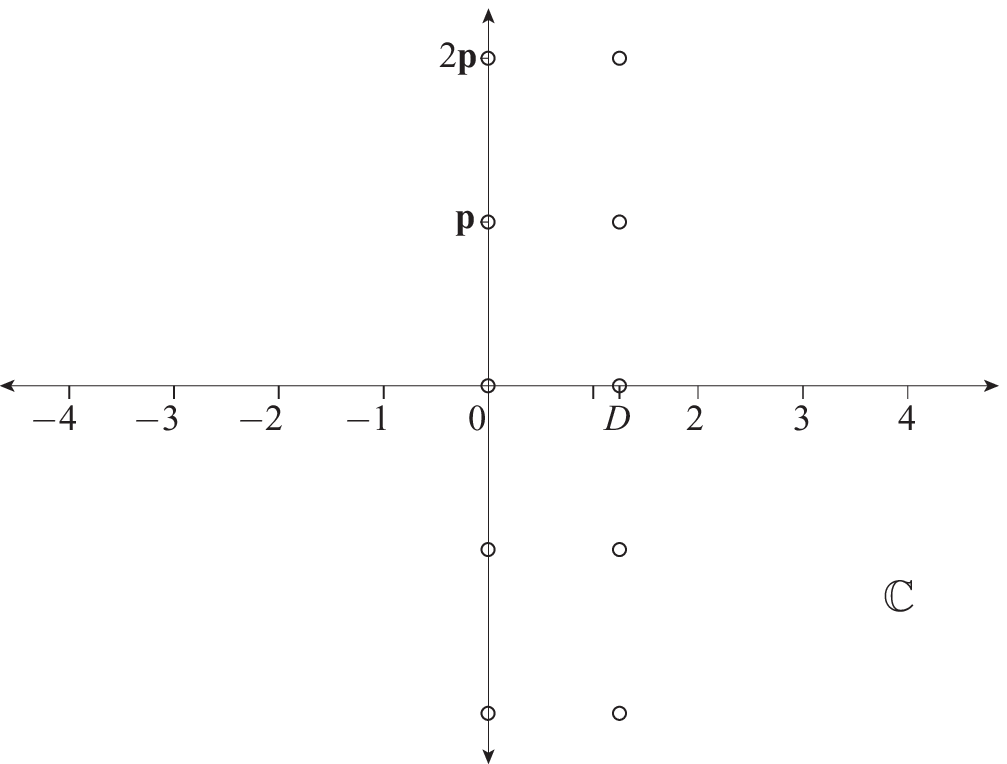}}
    \caption{The \possible complex dimensions of $K$ and $\del\gW$. The Minkowski dimension
      is $D=\log_34$ and the oscillatory period is $\per=\tfrac{2\gp}{\log3}$.}
    \label{fig:complex-dimensions} \centering
  \end{figure}

  The reader may easily check that formula \eqref{eqn:final-result}
  can also be written
  \begin{equation}
    \label{eqn:unproven corollary}
    V(\ge) = \sum_{n \in \bZ} \gf_n \ge^{2-D-in\per} + \sum_{n \in \bZ} \gy_n \ge^{2-in\per},
  \end{equation}
  for suitable constants $\gf_n, \gy_n$ which depend only on $n$.
  Now by analogy with the tube formula (\ref{eqn:explicit-tubular-formula-for-strings})
  from \cite{La-vF1}, we interpret the exponents of \ge in
  \eqref{eqn:unproven corollary} as the `complex co-dimensions' of \del\gW.
  Hence, we can simply read off the \possible complex dimensions, and as
  depicted in Fig.~\ref{fig:complex-dimensions}, we obtain a set of
  \possible complex dimensions
  \begin{equation}
    \label{eqn:poles by the picture}
    \D_{\del\gW} = \{D+in\per \suth n \in \bZ\} \cup \{in\per \suth n \in \bZ\}.
  \end{equation}

  One caveat should be mentioned: this is assuming that none of
  the coefficients $\gf_n$ or $\gy_n$ vanishes in
  \eqref{eqn:unproven corollary}. Indeed, in that case we would
  only be able to say that the complex dimensions are a subset
  of the right-hand side of \eqref{eqn:poles by the picture}.
  We expect that, following the `approximate tube formula'
  obtained in \cite{La-vF1}, the set of complex dimensions should
  contain all numbers of the form $D+in\per$.

  \begin{remark}
    \emph{
      We expect our methods to work for all lattice self-similar fractals
      (i.e., those for which the underlying scaling ratios are rationally dependent)
      as well as for other examples considered in [La-vF1] including the
      Cantor--Lebesgue curve, a self-affine fractal.
      For example, we have already
      obtained the counterpart of Thm.~\ref{thm:final-result} for
      the square (rather than triangular) snowflake curve.
      By applying density arguments (as in [La-vF1, Chap. 2]), our
      methods may also yield information about the complex dimensions of
      nonlattice fractals.
      }
  \end{remark}

\begin{remark}
    \label{comment:cor-to-main-nondirect}
    \emph{
    It should be pointed out that in the present paper, we
    do not provide a direct \dfn of the complex dimensions of the
    Koch snowflake curve \del\gW (or of other fractals in
    $\bR^2$). Instead, we reason by analogy with formula
    (\ref{eqn:explicit-tubular-formula-for-strings}) above to
    deduce from our tube formula (\ref{eqn:final-result})
    the \possible complex dimensions of \del\gW (or of
    $K$). As is seen in the proof of Thm.~\ref{thm:final-result},
    the tube formula for the Koch curve $K$ is
    of the same form as for that of the snowflake curve \del\gW.
    It follows that
    $K$ and \del\gW have the same \possible complex dimensions (see
    Cor. \ref{cor:complex-dim-of-koch}).
    In subsequent work, we plan to define the geometric zeta \fn
    $\gz_{\del\gW} = \gz_{\del\gW}(s)$ in this context to actually
    deduce the complex dimensions of \del\gW directly as the poles
    of the \mero \cntn of $\gz_{\del\gW}$.
    }

    \emph{
    In fact, this project is well underway.
    A self-similar tiling defined in terms of an iterated \fn system
    on \bRd is constructed in \cite{Pe}. In \cite{LaPe2}, we use this
    tiling and aspects of geometric measure theory discussed in
    \cite{LaPe1} to define a zeta function whose poles give the complex
    dimensions directly. We then use the zeta \fn and complex
    dimensions to obtain an explicit distributional inner tube
    formula for self-similar systems in \bRd, analogous to
    \cite[Chap. 6]{La-vF1}. We stress that although
    Thm.~\ref{thm:final-result} was helpful in developing
    \cite{Pe} and [\citen{LaPe1}--2], it is not a consequence or corollary
    of this more recent work.
    }
\end{remark}

\begin{remark}
  \label{rem:curvature plans}
    \emph{
    In the long-term, by analogy with Hermann Weyl's tube
    formula \cite{We} for smooth Riemannian submanifolds (see
    \cite{Gr}),
    we would like to interpret the \coeffs $a_n, b_n, \gs_n,
    \gt_n$ of the tube formula (\ref{eqn:final-result}) in terms of an
    appropriate substitute of the `Weyl curvatures' in this
    context. See the corresponding discussion in \cite[\S6.1.1
    and \S10.5]{La-vF1} for fractal strings; also see \cite{La-vF3}.
    }

    \emph{
    This is a very difficult open problem and is still far from
    being resolved, even in the one-dimensional case of fractal
    strings. See \cite[\S6.1.1]{La-vF1} and [\citen{LaPe1}--2].
    }
\end{remark}

\par

\begin{remark}(Reality principle.)
  \emph{
  As is the case for the complex
  dimensions of self-similar strings (see \cite[Chap. 2]{La-vF1}
  and [\citen{La-vF2}--3]), the \emph{possible complex dimensions of
  \del\gW come in complex conjugate pairs, with attached complex
  conjugate coefficients}. Indeed, since $\cj{g_{\ga}}=g_{-\ga}$
  (see \S\ref{subsec:formulating-h}), a simple inspection of the
  formulas in \eqref{eqn:mean-coeffs} shows that for every
  $n \in \bN$,
  \begin{equation}
    \label{eqn:conjugacy-of-coefficients}
    \cj{a_n} = a_{-n},
    \cj{b_n} = b_{-n},
    \cj{\gs_n} = \gs_{-n},
    \text{ and }
    \cj{\gt_n} = \gt_{-n}.
  \end{equation}
  It follows that $a_0,b_0,\gs_0$, and $\gt_0$ are
  reals and that $G_1$ and $G_2$ in \eqref{eqngrp:the periodic fns}
  are \emph{real}-valued, in agreement with the
  fact that $V(\ge)$ represents an \emph{area}.
    }

\end{remark}

\section{The Cantor-like \fn $h(\ge)$.}
  \label{subsec:formulating-h}
  We close this paper by further discussing the nonlinear
  Cantor-like \fn
  \begin{equation}
    \label{eqn:h-as-eps-expan}
    h(\ge) = \sum_{\ga \in \bZ} g_\ga (-1)^\ga\ge^{-i\ga\per}
      = \sum_{\ga \in \bZ} g_\ga e^{2\gp i \ga x}
      = g(x),
  \end{equation}
  introduced in \eqref{eqn:prelim-fourier-expn-of-h}.
  Note that since $h$ is real-valued (in fact, $0 \leq h < \gm < 1$), we
  have $g_{-\ga} = \cj{g_\ga}$ for all $\ga \in \bZ$.
  Further, recall from \S\ref{subsec:counting-error-blocks}
  that in view of the self-similarity of $K$, $h(\ge)$ is
  multiplicatively periodic with period 3, i.e.,
  \(h(\ge) = h(\tfrac\ge3).\)
  Alternatively, \eqref{eqn:h-as-eps-expan} shows that it can be thought of as an additively periodic
  \fn of $x$ with period 1, i.e., $g(x) = g(x+1)$.
  By the geometric definition of $h(\ge)$, we see that it is
  continuous and even monotonic when restricted to one of its
  period intervals
  \(I_n := \left(3^{-n-3/2},3^{-n-1/2}\right].\)
  Since $h$ is of bounded variation, its Fourier series converges
  pointwise by \cite[Thm. II.8.1]{Zy} and its Fourier \coeffs
  \saty
  \begin{equation}
    \label{eqn:rate-of-decay-of-g-alpha}
    g_\ga = O(1/\ga), \text{ as } \ga \to \pm \iy
  \end{equation}
  by \cite[Thm. II.4.12]{Zy}, as discussed in \S4.

\begin{figure}
    \scalebox{0.85}{\includegraphics{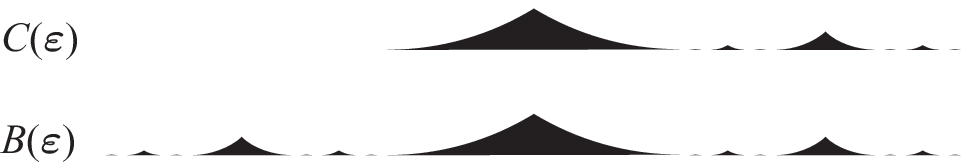}}
    \caption{\gm is the ratio $\vol[2](C(\ge)) / \vol[2](B(\ge))$.}
    \label{fig:describing_mu} \centering
  \end{figure}

  Further, since $h(\ge)$ is \defd as a ratio of areas (see
  Remark~\ref{rem:introduction of h}), we have
  $h(\ge) \in [0,\gm)$ for all $\ge > 0$. Note that $h(\ge) \leq \gm <
  1$ and so $h(\ge)$
  does not attain the value 1; the error blocks
  being formed are never complete. The partial error blocks only form across the
  first $\frac{2}{3}$ of the line segment beneath them. They reach
  this point precisely when $\ge = 3^{-n}/\sqrt{3}$ for some $n \geq 1$.
  Back in (\ref{eqn:error-block-area}), we found the error of a
  single error block to be given by
  \[B(\ge) = \sum_{k=1}^\iy 2^{k-1} A_k(\ge),\]
  where $A_k(\ge)$ is given by (\ref{eqn:kth-trianglet-area}).
  Thus the supremum of $h(\ge)$ will be the ratio
  \[\left(\tfrac{B(\ge_k)-A_1(\ge_k)}{2}+A_1(\ge_k)\right) / B(\ge_k),\]
  which will be the same constant for each
  $\ge_k = 3^{-k-1/2}$, \, $k=1,2,\dots$ \, (see
  Fig.~\ref{fig:describing_mu}).
  In other words, the number we need is
  \[\gm := \frac{A_1(\ge_k) + \tfrac{1}{2}\sum_{k=2}^\iy 2^{k-1} A_k(\ge_k)}{\sum_{k=1}^\iy 2^{k-1} A_k(\ge_k)}
         = \frac{A_1(\ge_k) + \sum_{k=2}^\iy 2^{k-2} A_k(\ge_k)}{\sum_{k=1}^\iy 2^{k-1} A_k(\ge_k)}
         \in (0,1).\]
  Note that although this definition of \gm initially appears to
  depend on $k$, the ratio in question is between two areas which
  have exactly the same proportion at each $\ge_k$; this is a
  direct consequence of the self-similarity of the Koch curve. In
  other words, if $C(\ge)$ is the area indicated in Fig.~\ref{fig:describing_mu}, then the relations
  \[3C(\tfrac{\ge}{3})=C(\ge) \text{ and } 3B(\tfrac{\ge}{3})=B(\ge)\]
  show that \gm is well-defined.

\begin{figure}
    \scalebox{0.80}{\includegraphics{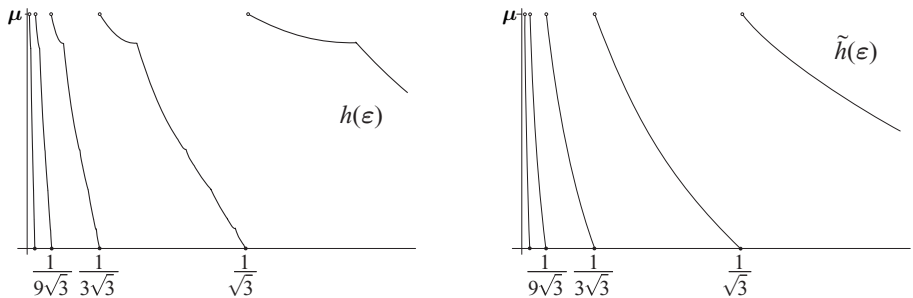}}
    \caption{A comparison between the graph of the Cantor-like \fn $h$
    and the graph of its approximation $\tilde{h}$.}
    \label{fig:h_vs_h} \centering
  \end{figure}

  We now approximate $h(\ge)$ by a function which
  shares its essential properties:
  \begin{Cond}
    \item $\displaystyle h(\ge_k)
           = \lim_{\vartheta \to 0^-}h(\ge_k+\vartheta) = 0$,
    \item $\displaystyle \lim_{\vartheta \to 0^+}h(\ge_k+\vartheta) = \gm$,
  \end{Cond}
  where for $k=1,2,\dots$, $\ge_k = \tfrac{3^{-k}}{\sqrt{3}}$ again.
  That is, $h(\ge)$ goes from 0 to \gm as \ge goes from $3^{-k}$ to
  $3^{-(k+1)}$. Using again the notation $x = -\log_3(\ge\sqrt3)$ and
  $\{x\} = x-[x]$, we see that the function
  \begin{equation}
    \label{eqn:defn-of-htilde}
    \htil = \gm \cdot \{-[x]-x\}
  \end{equation}
  shares both of these properties but is much smoother. Indeed, $\htil$ only
  has points of nondifferentiability at each $\ge_k$ and is
  otherwise a smooth logarithmic curve; see Fig.~\ref{fig:h_vs_h}.
  The true $h(\ge)$, by contrast, is a much more complex object that
  deserves further study in later work.

\newcommand{\bibgap}{0.1cm}
\newcommand{\bibart}[7]{#1, #2, \emph{#3} #4 \textbf{#5} (#6), #7.}

\newcommand{\bibbook}[5]{#1, \emph{#2}, #3, #4, #5.}

\newcommand{\bibvolume}[7]{#1, \emph{#2}, #3, vol. #4, #5, #6, #7.}

\newcommand{\bibbookp}[6]{#1, \emph{#2}, #3, #4, #5, pp.#6.}

\newcommand{\bibproc}[6]{#1, \emph{#2}, in: #3, #4, #5, pp. #6}
\newcommand{\bibprocvol}[2]{#1 (#2)} 
\newcommand{\bibprocinfo}[3]{#1, #2, #3}
\newcommand{\bibprocpub}[3]{#1, #2, #3}

\pgap

\par


\vstr

\par

\textsc{Michel L. Lapidus, \\{\scriptsize Department of
Mathematics, University of California, Riverside, CA 92521-0135}}
\par \emph{E-mail address:} \verb"lapidus@math.ucr.edu"

\vstr

\par

\textsc{Erin P. J. Pearse, \\{\scriptsize Department of
Mathematics, University of California, Riverside, CA 92521-0135}}
\par \emph{E-mail address:} \verb"erin@math.ucr.edu"

\end{document}